\documentstyle[12pt]{article}

\def\AJ{{\it Astrophys. J.} }
\def\ASP{{\it Astron. Soc. Pacific } }
\def\AJL{{\it Ap. J. Lett.} }

\def\CQG{{\it Class. Quantum Gravity} }

\def\GRG{{\it Gen. Relativity and Gravitation} }

\def\IJMP{{\it Int. J. Mod. Phys.} }

\def\NAT{{\it Nature} }

\def\NP{{\it Nucl. Phys.} }
\def\PL{{\it Phys. Lett.} }
\def\PR{{\it Phys. Rev.} }
\def\PRL{{\it Phys. Rev. Lett.} }
\def\PRTS{{\it Physics Reports} }

\def\De{\Delta}

\def\frac#1#2{{\textstyle{{#1}\over {#2}}}}
\def\ni{\noindent}
\def\lsim{\mathrel{\rlap{\lower4pt\hbox{\hskip1pt$\sim$}}
    \raise1pt\hbox{$<$}}}
\def\gsim{\mathrel{\rlap{\lower4pt\hbox{\hskip1pt$\sim$}}
    \raise1pt\hbox{$>$}}}
\def\sqr#1#2{{\vcenter{\vbox{\hrule height.#2pt
         \hbox{\vrule width.#2pt height#1pt \kern#1pt
         \vrule width.#2pt}
         \hrule height.#2pt}}}}

\newcommand{\beq}{\begin{equation}}
\newcommand{\eeq}{\end{equation}}
\newcommand{\bea}{\begin{eqnarray}}
\newcommand{\eea}{\end{eqnarray}}

\begin{document}
\titlepage

\begin{flushright}
{DF/IST-7.97\\}
{October 1998\\}
\end{flushright}
\vglue 2cm
	
\begin{center}
{{\bf Gamma-ray bursts, axion emission and string theory dilaton\\}
\vglue 1.5cm
{O.\ Bertolami \footnote{Also at Centro de F\'\i sica Nuclear da 
Universidade de Lisboa, Av. Gama Pinto 2, 1699 Lisboa Codex, Portugal.}\\}
\bigskip
{\it Instituto Superior T\'ecnico,
Departamento de F\'\i sica,\\}
\medskip
{\it Av.\ Rovisco Pais 1, 1096 Lisboa Codex, Portugal\\}
}
\vglue 2cm

\end{center}
\setlength{\baselineskip}{0.7cm}

\centerline{\bf  Abstract}
\vglue 1cm
\noindent
The emission of axions from supernovae is an interesting possibility 
to account for the Gamma-Ray Bursts 
provided their energy can be 
effectively converted into electromagnetic energy elsewhere. 
The connection between supernova and gamma-ray bursts 
has been recently confirmed by the observed correlation
between the burst of April 25, 1998 and the supernova SN1998bw.
We argue that the axion convertion into photons can be more 
efficient if one considers 
the coupling between an intermediate scale axion and the string 
theory dilaton along with the inclusion of string loops.
We also discuss the way dilaton dynamics may allow for a more 
effective energy exchange with electromagnetic radiation 
in the expansion process of fireballs.

\vfill
\newpage

\setcounter{equation}{0}
\setcounter{page}{2}
\baselineskip=20pt

\vglue 0.5cm
\section{Introduction}

Gamma-Ray Bursts are certainly among the most striking 
astrophysical discoveries of the 
century. Ever since their fortuitous discovery in late sixties 
\cite{klebesabel} these flashes of gamma radiation 
have been the subject of great interest and 
debate. There has been more than 3000 bursts recorded, about 2100 of which
by the Burst and Transient Source Experiment (BATSE) at Nasa's Compton 
Gamma-Ray Observatory \cite{meegan} and till February 1997 no
observation of bursts counterparts in any other wavelength had ever been 
recorded, although some evidence for an afterglow emission was found 
in high energy gamma rays when GEV photons were detected about $5 \times 
10^{3}$ seconds after the burst GRB 940217 \cite{hurley}. Data from 
BATSE clearly indicate that the 
distribution of bursts is isotropic around us although not homogeneous in a
3-dimensional Euclidean space \cite{fishman}. This last feature, which arises
from the lack of low intensity sources is, of course, a strong evidence for 
the cosmological nature of the bursts. This suspicion has been 
recently confirmed by the 
observations of the burst afterglows in the X-Ray region by 
BeppoSAX \cite{costa},
a dutch-italian satellite, and in the optical \cite{paradijs} for 
the burst of February 28th, 1997 (GRB 970228) and GRB 970508. For the
latter an afterglow in the radio has also been detected \cite{frail}.
These observations made it possible to determine the
afterglow's redshift, for instance, $z=0.835$ for GRB 970508 \cite{metzger}.
The identification of a host galaxy with $z = 3.42$ for the big burst of 
December 14, 1997 \cite{kulkarni1} provides further confirmation that
gamma-ray bursts are extragalactic in origin. Furthermore, the very recent
observation that the supernova SN1998bw located in the spiral
galaxy ESO 184 - G82 about $140$ milion light years away and the gamma-ray 
burst GRB980425 are related \cite{kulkarni2,galama,iwamoto} 
strongly suggests that at least some class
of gamma-ray bursts have its origin in supernovae. Actually, Colgate has
proposed the gamma-ray emission from supernovae as a possible origin of 
gamma-ray bursts more than twenty years ago \cite{colgate}, however it 
is only more recently that many of the theoretical problems ensued by 
this idea have been addressed.

The proposed models for explaining the GRBs observations 
require either quite drastic phenomena 
involving merging of compact objects and ultra relativistic motion 
of the source or new physics.
An example of the former is Paczy\'nski's hypernovae model 
which considers as a final result of some merging process
a $10~M_{\odot}$  Kerr black hole and the mechanism of 
Blandford-Znajek in order to obtain the extremely large magnetic fieds, 
$B \sim 10^{15}~G$, required to match the observed energy of the 
bursts, $E_{GRB} \sim 10^{52}~ergs$. A quite interesting
suggestion involving new physics assumes energy scapes from supernovae via
the emission of  
axions and is converted into electromagnetic energy elsewhere \cite{loeb}. 
This proposal solves the so-called compactness
problem related with the quick thermalization of the
electromagnetic energy that one expects in optically thick media due to
Compton scattering with electron-positron pairs \cite{goodman}. This is the 
usual situation one encounters in merging scenarios to explain 
gamma-ray bursts.
Furthermore, one has also to consider
that, in case the electromagnetic energy 
is contaminated by more than
$10^{-5}~M_{\odot}$ of baryons then it is shown that 
the photon energy tends to degrade
to lower energies implying that the burst duration will
be stretched beyond the observation limits \cite{eichler}. Neutron
star merging scenarios involve, for 
instance, about $10^{-2}~M_{\odot}$ of 
baryonic debris \cite{woosley}.  Of course, these difficulties can be, 
at least in principle, overcome assuming the source of the bursts is
accelerated to the ultra relativistic regime as can be achieved in
fireballs created from the merging of compact objects 
(see for instance Refs. \cite{meszaros} for recent reviews).   
The non-thermal spectra of the gamma-ray bursts suggest a sudden 
energy release as encountered
in some situations in cosmology such as for instance in bubble collisions in 
first order phase transitions or resonant decay of bosonic fields into
radiation.
A quick estimate however, shows that the simplest implementation of
these ideas is irrelevant in the gamma-ray bursts problem.
Indeed, consider for instance, that $\Omega_{V} < 0.5$ 
(which is supported by the
most recent studies of the cosmological parameters \cite{perlmutter}), then 
the available vacuum 
energy is about $\rho_{V} \sim 5 \times 10^{-6}~h_{0}^{2}~GeV~cm^{-3}$ 
($h_{0}$ being the Hubble constant in units $100~km~s^{-1}~Mpc^{-1}$) 
which is
just a tiny fraction, even when allowing for a large redshift factor,
of the energy density in gamma-ray bursts assuming the bursts are cosmological 
and that their source is fairly 
compact $D_{source} < 3 \times 10^{3}~km$ as can be estimated from smallest
observed time variability interval $\De T < 10~ms$ \cite{piran1}. 
In this work we consider the proposal of Ref. \cite{loeb} in
the context of string theory. As we discuss below,
in string theories the well known cosmological upper bound for the 
vacuum expectation value for the breaking of the Peccei-Quinn
symmetry, $f_{PQ} < 10^{12}~GeV$, may not hold due to the presence of 
the moduli fields, which on its own prevents axions as being the 
putative source of the gamma-ray bursts if axions are allowed in Grand Unified 
Theories (GUTs), even though the inclusion of string loops may 
allow, under conditions, $f_{PQ} \approx few~10^{12}~GeV$. 
These conclusions are, of course, strongly dependent on the estimates
of the energy release of gamma-ray bursts and may be relaxed if
$E_{GRB} < 10^{52}~ergs$.
We shall also point out that our scheme is compatible with the 
mini-supernova model of
Ref. \cite{blinnikov} provided the axion energy can be released nearby
the exterior layers of red giants. It
is worth stressing the naturalness of 
considering astrophysical and cosmological
phenomena involving axions since these particles  
stand out as the most prominent candidates 
for the dark matter of the Universe 
in the much observationally favoured Cold Dark Matter Models.
Finally, we shall see that independently of the presence of axions 
the dynamics of the string theory
dilaton may play a role in the fireball expansion process.

\section{Axion-Dilaton Coupling}

Axions arise in particle physics to explain the strong CP problem via
the Peccei-Quinn mechanism \cite{peccei} (see eg. Ref. \cite{kim} 
for a review). Axions are pseudogoldstone bosons
associated with the spontaneous breaking of an anomalous chiral
$U(1)_{PQ}$ symmetry with vacuum expectation value, $f_{PQ}$. 
Axions acquire mass as QCD instantons break $U(1)_{PQ}$ and 
$m_{a} = {f_{\pi}~m_{\pi} \over f_{PQ}}$, where $f_{\pi} = 93~MeV$
and $m_{\pi} = 134.5~MeV$. For standard axions, bounds on $f_{PQ}$
can be obtained from astrophysical and cosmological considerations \cite{kim}:
\beq
10^{9}~GeV < f_{PQ} < 10^{12}~GeV 
\quad ,
\label{fpq}
\eeq
where the lower bound arises from the demand that axions do not lead 
to a quick cooling of stars while the upper bound comes from requiring that
the classical coherent oscillations of axions do not dominate the Universe
dynamics at present \cite{preskill}. In the context of string theory, 
however, the cosmological bound may not hold as moduli fields
which are assumed to have masses comparable to the gravitino  
are substantially heavier than the axions, implying that 
their coherent oscillations dominate the Universe dynamics 
quite early on \cite{dvali}.  This may imply that the axion can be much 
more ``invisible'' than usually assumed and that 
axions arising from GUTs or supersymmetric GUTs 
via the breaking of R-type symmetries are cosmologically acceptable. 
The possibility that $f_{PQ}$ is greater than the intermediate scale 
$10^{12}~GeV$ renders   
axions unfeasible as candidates for solving the compactness 
problem of gamma-ray bursts if one assumes that the energy release
of gamma-ray bursts is indeed $E_{GRB} \sim 10^{52}~ergs$. 
We shall see that this conclusion  
remains essentially unaltered if one considers 
the effect of string loops and assumes these can stabilize the
dilaton in the absence of a potential \cite{damour}. This is a 
quite relevant suggestion as moduli
fields, including the dilaton, are shown to remain massles in all orders in
string perturbation theory. Indeed, after assuming that 
the dilaton coupling is 
universal, the string dynamics at low energies compared with the Planck
mass, $M_P$, is described by
the following lowest order bosonic action involving axion, dilaton, gravity
and Yang-Mills fields \cite{damour}:
\beq
S_B=\int d^4x \sqrt{-g} \left\{ -{R \over 2 k^2} + 2 (\partial
\phi)^2 + {k_{i} \over 4}~B(\phi)~F_{\mu\nu}^a F^{\mu\nu a} + 
{b_{i} \over 4}~B(\phi)^{3}~{a \over f_{PQ}}  
F_{\mu\nu}^a\tilde F^{\mu\nu a} + ... \right\}
\quad ,
\label{action}
\eeq
\ni
where $k^2=8\pi M_P^{-2}$, $k_{i}$ and $b_{i}$ 
are order one constants, 
the field strength $F_{\mu\nu}^a$ corresponds to 
the one of a Yang-Mills theory with gauge group $G_{i}$ which is 
a subgroup of $E_8\times E_8$ or $SO(32)$
and $\tilde F^{\mu\nu}$ is the dual of the gauge field strength;  
moreover, following 
Ref. \cite{damour}, we introduce the universal function 
of the dilaton, $\phi$: 
\beq
B(\phi) = e^{- 2 k \phi} + c_0 + c_1 e^{2 k \phi} + c_2 e^{4 k \phi}
+ ...  
\quad ,
\label{bfunction}
\eeq
\ni
which expresses the fact that string-loop interactions
have an expansion in powers of the dilaton, 
that is $g_S \equiv e^{2 k \phi}$; the coefficients
$c_0, c_1, c_2$, ... are presently unknown. As discussed in 
\cite{damour},  when accounting for string loops the dynamics of 
the dilaton is such that the function $B(\phi)$ must reach a 
maximum at present, when say $\phi = \phi_{0}$, as fermion masses 
are shown to be proportional to inverse powers of $B(\phi)$.
The gauge coupling constants are extracted from the dilaton-gauge field 
coupling  ${k_{i} \over 4}~B(\phi)~Tr\left(F_{\mu\nu} F^{\mu\nu}\right)$ 
in the bosonic action,
from which follows that $g_{i}^{-2} = k_{i}~B(\phi_{0})$.
Of course, the coupling between the axion and 
the electromagnetic field, i.e. $-{c_{a \gamma \gamma} \over 2}\alpha_{EM}
B(\phi)^{4}{a \over f_{PQ}} {\bf E.B}$, is contained 
in the third term in (2) for the $U_{EM}(1)$ subgroup of $G$ such that 
$c_{a \gamma \gamma}$ hides all the gauge symmetry branching from
the GUT gauge group down to $U_{EM}(1)$. For the simplest case
$c_{a \gamma \gamma} = 8\pi^{2}k_{1}b_{1}$. Notice that we have not
included the function $B(\phi)$ in the definition of $c_{a \gamma \gamma}$. 
Given the universality of the dilaton coupling, the quark sector
has a generic term ${c_{af} \over 2 f_{PQ}}
B(\phi) {\bar \psi} \gamma_{\mu} \gamma_5 \psi \partial^{\mu} a$
\cite{raffelt}. 
The coupling between the axion and the gauge field 
ensures the photon conversion of axions due 
to magnetic fields surrounding the progenitor neutron star or interstellar 
medium. The conversion of axions (and also omions, the Nambu-Goldstone 
bosons outside the core of global cosmic strings) into photons 
has been discussed in Refs. \cite{sikivie} with the conclusion 
that strong inhomogeneous magnetic fields are required and that 
the conversion probability is proportional $\vert {\bf B} \vert^2$
and is suppressed by a factor 
$(\omega_{pl}^{2} - m_{a}^2)^{-1/2}$ where ${\bf B}$ is the 
magnetic field and $\omega_{pl}$ the plasma 
frequency which is an effective mass for the photon. Of course, the conversion
of axions is greatly enhanced if $ \omega_{pl} \approx m_a$. We can hence 
conclude that an efficient conversion of axions into photons requires strong
inhomogeneous magnetic fields and media where the plasma frequency is close to 
the axion mass. The difficluty in meeting the conditions for 
efficient conversion into photons may be at the origin of 
the difference of rates between supernova and gamma-ray bursts. 

For $m_{a} << 10^{-2}~eV$ axions free stream through neutron stars
with temperature $T_{NS}$ leading, for standard axions, to a total luminosity
given by \cite{raffelt}:
\beq
L_{a} \approx 2~\times 10^{50}~\times10^{\pm 1.5}
~\left({m_a \over 10^{-4}~eV}\right)^2
~\left({T_{NS} \over 30~MeV}\right)^{3.5}~erg~s^{-1}~.
\label{lumin}
\eeq

\noindent 
This luminosity is on its own inconsistent, 
for $m_{a} \approx 10^{-4}~eV$, with the best-fit  
gamma-ray burst standard candle luminosity, 
$L_{GRB} \sim 1.6 \times 10^{52}~erg~s^{-1}$, 
for $\Omega = 1$ and $H_{0} = 50~km~s^{-1}~Mpc^{-1}$ \cite{wijers} and the
possibility of absence in string theory of an upper 
bound for $f_{PQ}$ makes things 
worse as axions may be even lighter than $10^{-5}~eV$. 
The inclusion of string loops alters slightly this as one should then 
multiply the luminosity (eq. (4)) by a factor $B(\phi_{0})^{2}$ 
and $k_{i}B(\phi_{0}) \sim 10$. Thus, $L_{GRB} \approx 10^{52}~erg~s^{-1}$
can be obtained for $f_{PQ} \approx 10^{11}~GeV$. Notice that,
for instance, the SN1998bw 
explosion can be modelled by a spherically symmetric explosion of a 
quite massive star with a total
ejected mass of about $10~M_{\odot}$ and approximately $3 \times 10^{52}~ergs$
of kinetic energy \cite{galama}, although it has been argued 
that energy releases of this magnitude actually reflect
collimation effects and Doppler boosting \cite{wang}. This collimation
implies that the energy output of gamma-ray bursts may be actually
much smaller, $E_{GRB} \sim 10^{46} - 10^{48}~ergs$, \cite{wang} and
suggests a higher rate of bursts possibly consistent with the
supernovae rate \cite{wang,cen}. 
Energy considerations can also be relaxed if one considers the mini-supernova
model proposed by Blinnikov and Postnov \cite{blinnikov} in which an 
optical afterglow can be produced by the release of about $10^{50}~ergs$
nearby the exterior $10^{-3}~M_{\odot}$ layers of a red giant. This energy 
can be achieved by the emission of axions by supernovae and their conversion 
into radiation due to the strong magnetic fields in the outer
layers of the red giant provided that
$f_{PQ} \approx 10^{12}~GeV$ if string loops are taken into account.
Hence, we see that string loop effects can help the case of standard axions
as mediators of the supernova burst energy, but on no ways they can save the 
case of axions arising from GUTs, unless the energy release of GRBs is much
smaller than $10^{52}~ergs$.

\section{Fireball Expansion}

Another way the dilaton may play a role is in the fireball expansion 
after the merging of compact objects,
specially if one assumes, as suggested from the observation of 
GRB970228, GRB970508, GRB970828 and  GRB971214, 
that the bursts afterglows are close to
star forming regions with rather high redshifts, say $z \gsim 1$ 
\cite{paczynski1}.
Supposing that the expansion of the fireball and the ensued 
cooling is adiabatic, then the process 
can be described through the Milne cosmological model \cite{piran2}. 
If one further
assumes the radiation corresponds, for simplicity, to a 
triplet of massive vector fields with an $SO_{I}(3)$ internal global symmetry
\cite{bento1}, it follows that one can build an homogeneous 
and isotropic model and show 
that the dilaton energy can be transferred into radiation as 
far as $B(\phi) > 0$ and its derivative with respect to $\phi$ is negative.
This energy transfer has been previously discussed in 
the cosmological setting related with 
the so-called Polonyi problem \cite{bento2} for the case of an 
$SO(3)$ gauge field  \cite{bertolami}. 
  
Indeed, considering only  homogeneous and isotropic
field configurations on a spatially flat spacetime, the most general
metric is given in terms of the lapse function, 
N(t), and the scale factor, R(t): 
\beq
ds^2=-N^2(t) dt^2 + R^2(t) d\Omega^2_3 ~.
\quad
\label{metrica}
\eeq
\ni
We consider for simplicity  a 
triplet of massive vector fields with an $SO_{I}(3)$ global symmetry,
our conclusions however, are qualitatively independent of
this choice. We then use the following homogeneous and 
isotropic Ansatz for the vector field \cite{bento1}:
\beq
A_0 = 0~~;~~~A_i (t) dx^i = \sum_{i=1}^{3} \chi_0(t) L_{i}dx^i
\qquad
\label{gauge}
\eeq
\ni
$\chi_0(t) $ being an arbitrary function of time and $L_i$ the generators
of $SO_{I}(3)$.

We start by dimensionally reducing action (\ref{action}), allowing
only for homogeneous and isotropic field configurations and 
dropping the axion term. 
This procedure allows treating the
contribution of the vector fields on the same footing as the
remaining fields, as opposed to the usual treatment of radiation as a
fluid.

Introducing the Ans\"atze (\ref{metrica}) and (\ref{gauge}) into the action 
(\ref{action}) leads after introducing a mass term for the vector field 
to the following {\it effective} action for 
the dilaton-Einstein-Proca system after
integration over $\bf {R}^3$ and division by the infinite volume of its
orbits:
\beq
S_{eff}=-\int_{t_1}^{t_2} dt \left\{ -{3 {\dot R}^2 R \over k^2 N}
+ {3 R\over N} B(\phi) \left[ {{\dot\chi_0}^2\over 2} - N^2 m^2\chi_0^2 
- {N^2 \over R^2} {\chi_0^4\over 8}\right] 
+ { 2 R^3\over N} {\dot \phi}^2\right\},
\qquad
\label{effective}
\eeq
\ni
where the dots denote time derivatives and $m$ the mass of the vector fields.
The equations of motion in 
the N=1 gauge are given by:
\beq
2 {\ddot R \over R} + H^2 + {k^2 \over 3} B(\phi)
\rho_{\chi_0} + 2 k^2 {\dot \phi}^2 = 0~,  
\qquad
\label{a}
\eeq

\beq
\ddot \phi + 3 H {\dot \phi} - {1\over 4} {B^\prime(\phi)} \zeta_{\chi_o} 
 = 0~
\qquad
\label{phi}
\eeq

\beq
{\ddot \chi_0} + [H+{B^\prime(\phi)\over B(\phi)}  \dot \phi] {\dot \chi_0} 
+ 2 m^2 \chi_0 + {\chi_0^3 \over 2 R^2} =0~,
\qquad
\label{chi}
\eeq
\ni
where the primes denote derivatives with respect to $\phi$, $H=\dot R / R$ 
is the fireball rate of expansion, 
$\rho_{\chi_o}  = 3  \left[{{\dot \chi_0}^2\over 2 R^2} 
+ {m^2 \chi_0^2 \over R^2} + {\chi_0^4 \over 8 R^4} \right]$ and
$\zeta_{\chi_0} =  3  \left[{{\dot \chi}_0^2\over 2 R^2} 
- {m^2 \chi_0^2 \over R^2} - {\chi_0^4 \over 8 R^4} \right]$.

Furthermore, the Friedmann equation is obtained extremizing the effective
action (\ref{effective}) with respect to the lapse function:
\beq
H^2 = {k^2\over 3} \left[ 4 \rho_\phi + B(\phi)\rho_{\chi_o}\right]~, 
\qquad
\label{friedmann}
\eeq         
\ni
where $\rho_\phi = {1\over 2} {\dot \phi}^2$.

\ni
Our field treatment of radiation reveals an energy exchange mechanism 
that may turn out to be relevant in the process of expansion 
of the fireball. 
Working out the equations above, 
we obtain the energy exchange equations: 
\beq
{\dot \rho}_{\phi} = - 3 H  {\dot
\phi}^2 +  {1\over 4} {B^\prime(\phi)}  \zeta_{\chi_o}{\dot \phi}~,
\qquad
\label{rhophi}
\eeq

\beq
{\dot \rho}_{\chi_0} = -4 H \rho_{\chi_0} -  3
{B^\prime(\phi)\over B(\phi)} {\dot \chi_0^2\over R^2} \dot \phi ~. 
\qquad
\label{rhochi}
\eeq

\ni
The new feature in these equations are the
terms proportional to $\dot \phi$. If gamma-ray bursts are prone to occur
in star forming regions with rather high redshifts, then the dynamics of the 
dilaton field may be relevant. Indeed, if $B(\phi) > 0$ and its 
derivative with respect to $\phi$ is negative then we see that radiation 
acquires the energy lost by the dilaton despite losses due to the fireball
expansion depicted by terms proportional to $H$. 
Naturally, the energy exchange becomes less and less efficient as the fireball
expands and it occurs predominantly when $H \approx 
{B^\prime(\phi)\over B(\phi)} \dot \phi \approx m$.  
This situation is similar to the one encountered in cosmology
after inflation but prior the reheating phase \cite{bento2}. 
Of course, as the photon in a plasma medium behaves as if having a 
mass $\omega_{pl}$, then we can identify our triplet of massive vector
fields with the electromagentic field and therefore 
$m = \omega_{pl}$. For small $\omega_{pl}$ and $\dot \phi$ our mechanism 
remains effective for quite a long time, provided the discussed conditions for 
$B(\phi)$ are satisfied. Even considering that, at a fundamental level, 
the electromagnetic radiation cannot be treated, due to its lack 
of rotational symmetry, as performed bove, 
the basic features of the energy exchange mechanism are still present.
This can be
seen from the energy-momentum conservation equation
\beq
T_{\mu\nu~;\mu}^{EM} = - B(\phi)^{-1} \left[T_{\mu\nu;\mu}^{\phi} 
+ B(\phi)_{;\mu} T_{\mu\nu}^{EM}\right]~,
\qquad
\label{emtensor}
\eeq
which shows that a liquid transfer of dilaton energy into radiation may 
occur if $B(\phi)_{;\mu} < 0$. 
Nevertheless, it is worth emphasizing that 
our treatment of radiation as a triplet of 
massive vector fields seems to be more appropriate in capturing
the main features of the energy exchange between fields in a medium.

\section{Discussion and Conclusions}

Gamma-ray bursts may possibly be the sole astrophysical 
evidence we have encountered so far suggesting new physics beyond the
standard model. In this respect, massive objects of pure quark matter 
\cite{anoushirvani}, quantum gravity effects in the propagation of 
electromagnetic waves in vacuo \cite{amelino-camelia}, 
neutron star explosion caused by accumulation of Q-balls \cite{kusenko}, 
to mention just a few
have been suggested in connection with these bursts. In any case, 
gamma-ray bursts are striking phenomena requiring  
extreme astrophysical conditions. 
In this work  we have argued 
that axions emitted from supernovae, as suggested in \cite{loeb},
can be a potential explanation for the origin of the
gamma-ray bursts. This possibility, although far from being proven, has 
gained support from the identification
of the supernova SN1988bw as the source of the burst GRB980425.
We have shown that, in the context of string theory, string loops may play 
a role in achieving the axion energy required to match the observations, 
for $f_{PQ} \lsim 10^{11}~GeV$ if $E_{GRB} \sim 10^{52}~ergs$. 
For $f_{PQ} \approx 10^{12}~GeV$
we envisage that only via the presence of red giants afterglows 
can be explained, as suggested in \cite{blinnikov}. This points to a distinct
observational signature, namely the association of afterglows with supernovae
{\it and} red giants. The large magnetic fields present in
the outer layers of red giants may quite possibly be effective in the  
conversion into radiation of the energy from supernovae 
carried away by axions. Difficulty in satisfying the resonance condition, 
$\omega_{pl} \approx m_a$, and the small probability of association
between supernovae and nearby red giants arise as an elegant explanation for
the different rates of supernovae and gamma-ray bursts.
In this context, one could also consider 
the possibility that the dilaton may acquire a potential 
(and hence a mass) non-perturbatively, as usually discussed, via the 
process of 
condensation of gauginos. Thus, a possible scenario would be accounting 
for the disruption of red giants outer 
layers due to the dilaton decay. For that, one should require as a necessary
condition that the dilatons would live at least as long as about the age of the
Universe, that is 
$\tau_{\phi} \equiv \Gamma_{\phi}^{-1} = M_{P}^2/8 \pi m_{\phi}^{3} 
\approx 10^{17}s$, from which follows that $m_{\phi} \sim 25~eV$. 
Notice that for this 
purpose the radiative decay of the axion 
is totally irrelevant as its lifetime is about 
$10^{45}~(m_a/10^{-4}~eV)^{-5}~s$.

Finally, we have shown that independently of the presence of
axions, the dilaton dynamics may itself be relevant 
in the fireball expansion process in scenarios involving merging of compact 
objects and we have set, through the analysis of the energy exchange  
between the dilaton and a triplet of massive vector fields in an
homogeneous, isotropic and adiabatic model, the conditions under 
which this dynamics is important, namely when 
$B(\phi) > 0$ and  $B(\phi)_{;\mu} < 0$.

\end{document}